\documentclass{article}
\usepackage[english]{babel}
 
\usepackage{hyperref}
\hypersetup{
    colorlinks=true,
    citecolor=blue,
    linkcolor=blue,
    filecolor=magenta,      
    urlcolor=cyan,
}
\usepackage{graphicx,epsfig,color,xcolor}
\usepackage[english]{babel}
\usepackage{amssymb,amsfonts,amsmath}
\usepackage{verbatim}
\usepackage{bbm,bbold}
\usepackage{cite}
\newcommand{\be}{\begin{equation}} \newcommand{\ee}{\end{equation}}
\newcommand{\bea}{\begin{eqnarray}} \newcommand{\eea}{\end{eqnarray}}
\newcommand{\beann}{\begin{eqnarray*}}  \newcommand{\eeann}{\end{eqnarray*}}
\newcommand{\bfig}{\begin{figure}} \newcommand{\efig}{\end{figure}}
\newcommand{\ba}{\begin{array}} \newcommand{\ea}{\end{array}}
\newcommand{\bcen}{\begin{center}} \newcommand{\ecen}{\end{center}}
\newcommand{\btab}{\begin{tabular}} \newcommand{\etab}{\end{tabular}}

\newtheorem{Proposition}{Proposition}[section]

\newtheorem{Theorem}{Theorem}[section]
\newtheorem{Lemma}{Lemma}[section]
\newtheorem{Corrolary}{Corrolary}[section]

\newcommand{\bp}{\begin{Proposition}}   \newcommand{\ep}{\end{Proposition}}
\newcommand{\bt}{\begin{Theorem}}   \newcommand{\et}{\end{Theorem}}
\newcommand{\bl}{\begin{Lemma}}     \newcommand{\el}{\end{Lemma}}
\newcommand{\bc}{\begin{Corrolary}} \newcommand{\ec}{\end{Corrolary}}

\def\Xint#1{\mathchoice
{\XXint\displaystyle\textstyle{#1}}%
{\XXint\textstyle\scriptstyle{#1}}%
{\XXint\scriptstyle\scriptscriptstyle{#1}}%
{\XXint\scriptscriptstyle\scriptscriptstyle{#1}}%
\!\int}
\def\XXint#1#2#3{{\setbox0=\hbox{$#1{#2#3}{\int}$}
\vcenter{\hbox{$#2#3$}}\kern-.5\wd0}}

\def\dashint{\Xint-}

\begin{document}

\begin{center}

\centering{\Large {\bf Holographic RG flow and reparametrization invariance of Wilson loops}}

\vspace{8mm}

\renewcommand\thefootnote{\mbox{$\fnsymbol{footnote}$}}
Diego Gutiez${}^{1,2}$\footnote{gutiezdiego@uniovi.es} 
Carlos Hoyos,${}^{1,2}$\footnote{hoyoscarlos@uniovi.es}

\vspace{4mm}
${}^1${\small \sl Department of Physics, Universidad de Oviedo} \\
{\small \sl c/ Leopoldo Calvo Sotelo, ES-33007 Oviedo, Spain} 

\vspace{2mm}
\vskip 0.2cm
${}^2${\small \sl Instituto Universitario de Ciencias y Tecnolog\'{\i}as Espaciales de Asturias (ICTEA)}\\
{\small \sl  Calle de la Independencia, 13, 33004 Oviedo, Spain}

\end{center}

\vspace{0mm}

\renewcommand\thefootnote{\mbox{\arabic{footnote}}}

\begin{abstract}
\noindent  

We study the fate of reparametrization invariance of Wilson loops, also known as 'zig-zag' symmetry, under the RG flow using some simple cases as guidance. We restrict our analysis to large-$N$, strongly coupled CFTs and use the holographic dual  description of a Wilson loop as a fundamental string embedded in asymptotically AdS spaces, at zero and nonzero temperature. We then introduce a cutoff in the holographic radial direction and integrate out the the section of the string closer to the AdS boundary in the spirit of holographic Wilsonian renormalization. We make explicit the map between Wilson loop reparametrizations and conformal transformation of the string worldsheet and show that a cutoff  anchored to the worldsheet breaks conformal invariance and induces an effective defect action for reparametrizations at the cutoff scale, in a way similar to nearly-$AdS_2$ gravity or SYK models. On the other hand, a cutoff in the target space breaks worldsheet diffeomorphisms and Weyl transformations but keeps conformal transformations unbroken and does not generate a non-trivial action for reparametrizations. 

\end{abstract}

\thispagestyle{empty}

\newpage

\tableofcontents



\section{Introduction}

\begin{figure}[h!t]
\begin{center}
\includegraphics[width=6cm]{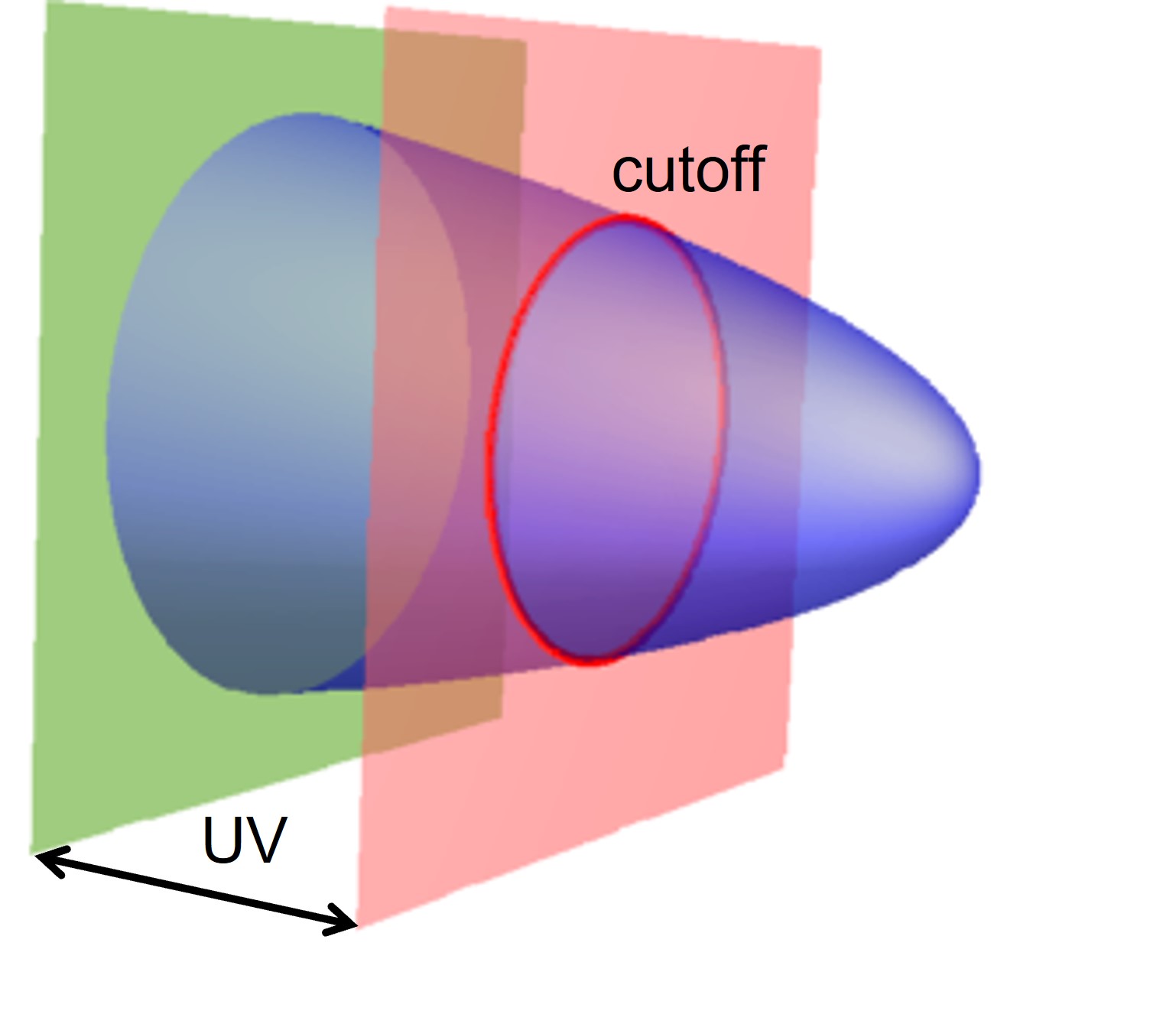}
\caption{\small The string dual to a Wilson loop is attached to a circle at the boundary (back green plane) and extends along the holographic radial direction. Introducing a radial cutoff (front red plane) separates the geometry in the UV region between the cutoff plane and the boundary, and the IR region at the other side of the cutoff. Removing the UV region leaves the IR section of the string plus a cutoff action defined at the boundary of the truncated string (red circle).}\label{fig:cutoff}
\end{center}
\end{figure}

Gauge/gravity duality has been extensively used as a phenomenological tool to describe strongly coupled systems in particle physics and condensed matter (see e.g. \cite{Hartnoll:2009sz,Brambilla:2014jmp,DeWolfe:2018dkl}). In the search of holographic duals of realistic theories like QCD one has to face the problem that the UV physics cannot be captured in general by a weakly coupled gravity dual, making the problem effectively intractable in this regime. Provided the identification between the energy scale in the field theory side and the position along the holographic radial direction in the dual, a possible way out of this issue is to introduce a radial cutoff in the gravity side, thus dividing the geometry in an ``IR'' region on one side of the cutoff and a ``UV'' region on the other side, as in Figure~\ref{fig:cutoff}. Once this is done one can give away with the troublesome UV region and work only with the IR region where the gravitational theory is weakly coupled. An important problem with this approach is that observables in the field theory have to be read from the asymptotic behavior of the fields in the UV region. Then, in order to be able to extract any useful information from the gravity dual, one needs to introduce a prescription that allows to recover the UV information after the UV region has been removed.

Some intuition can be gained from studying this problem in a gravity dual that is weakly coupled everywhere. One can then introduce the radial cutoff and see what is necessary in order to reproduce the same values for the observables one would have been obtained from the full geometry. Since in practice one is solving classical equations of motion for the gravity fields, this amounts in the end to determining the boundary conditions for the fields at the radial cutoff. This can be accomplished by introducing an effective action at the cutoff. The cutoff effective action can be though of as an effective action in the field theory, obtained after integrating out UV degrees of freedom, and coupled to the strongly coupled sector described by the IR region of the gravity dual. The cutoff action depends on the values of the gravitational fields and their derivatives at the cutoff. The cutoff action follows the general rules of effective field theories, it is constrained by symmetries, admits an expansion in small derivatives and all the information about the UV region is hidden in the values of the coefficients appearing in the action. Then, by fitting the coefficients of the cutoff action, one may be able to capture the right UV physics in the holographic dual, while remaining in the region where gravity is weakly coupled. For Wilson loops, which is our topic of interest here, the cutoff action has been studied in \cite{Casalderrey-Solana:2019vnc,Gutiez:2020sxg}, following the general approach of holographic Wilsonian renormalization \cite{Heemskerk:2010hk,Faulkner:2010jy}.

Wilson loops provide a complete set of gauge invariant operators that could in principle be used to compute any observable in Yang-Mills theory, and determine directly some important phenomenological quantities like the quark-antiquark potential. Their expectation value in the large-$N$ limit can be computed at strong coupling by means of the gauge/gravity duality. For a CFT like ${\cal N}=4$ super Yang-Mills, the holographic dual of the Wilson loop is a fundamental string anchored at the asymptotic $AdS$ boundary of the gravity dual geometry \cite{Maldacena:1998im}.

Wilson loops enjoy a quite large reparametrization invariance. By their definition, they are determined by the holonomy of the gauge field along a closed curve ${\cal C}$ \footnote{We will consider a Wilson loop extending to spatial infinity as ``closed'', from the point of view that gauge transformations should go to a constant at infinity in flat space.}. The curve itself can be parametrized as the trayectory of a particle $x^\mu(\tau)$, where $\tau$ parametrizes the wordline, in such a way that
\begin{equation}
\oint_{\cal C} A=\int d\tau \, \dot{x}^\mu A_\mu[x(\tau)].
\end{equation}
Any parametrization whose image is said curve should lead to the same value for the Wilson loop, even if it traces the curve back and forth, this is the so-called ``zig-zag'' symmetry \cite{Polyakov:1997tj}. This is true at strong coupling even for 1/2 BPS Wilson loops as noted in \cite{Drukker:1999zq}, even though the coupling to the scalar fields in principle breaks the zig-zag symmetry. The difference in the holographic dual between ordinary and BPS Wilson loops are boundary conditions for the string along the internal space, but the configurations we will study are valid for both sets of boundary conditions, so the symmetries turn out to be the same in this case.

It is unclear whether introducing a cutoff as described above preserves or breaks reparametrization invariance of the Wilson loop, in the same way that other symmetries such as conformal invariance are broken. If that were the case, the cutoff action should include additional terms that compensate the non-invariance of the string in the IR region, in a way analogous to the anomaly inflow mechanism of gauge theories. Clarifying this point will be our goal in this work. In order to proceed, it will be much more illuminating to work with the Polyakov action for the string dual to the Wilson loop, rather than the Nambu-Goto action that it is usually employed. As we will see, in the gravity dual description the reparametrization invariance of the Wilson loop corresponds to the conformal invariance of the string. This is the analog of the usual map between isometries of the geometry in the gravity side and global symmetries of the field theory dual, except in this case we are treating with a group with an infinite number of generators.

It turns out that the fate of reparametrization invariance when a radial cutoff is introduced depends on how this is done. If the radial cutoff is on the worldsheet the situation is similar to the nearly-$AdS_2$ physics described in \cite{Maldacena:2016upp}, connected to Jackiw-Teitelboim (JT) gravity \cite{Jackiw:1984je,Teitelboim:1983ux,Almheiri:2014cka} and the Sachdev-Ye-Kitaev model (SYK) \cite{Sachdev:1993,Kitaev:2015,Polchinski:2016xgd,Maldacena:2016hyu} (see \cite{Sarosi:2017ykf} for a review on these topics). Effectively reparametrization invariance is broken and there is an effective action at the cutoff proportional to a Schwarzian derivative of the Goldstones associated to the broken symmetries. On the other hand, if the radial cutoff is at a fixed position on the target space of the string, reparametrization invariance is unbroken. In this second case both worldsheet diffeomorphisms and Weyl transformations are broken by the cutoff, so there are Schwarzian effective actions for the broken (gauge) symmetries, but they cancel out when a conformal transformation involving both is considered.

The content of the paper is as follows. In section \ref{sec:line} we first study in quite detail a straight Wilson line in a CFT, described by a fundamental string in the holographic dual with Polyakov action. We introduce a radial cutoff and derive the cutoff action by integrating the section of the string in the UV region. We then discuss in detail the reparametrization symmetries and the cutoff action. Next, in section \ref{sec:gener}, we generalize the results to other simple cases in a CFT: a circular Wilson loop at zero temperature and the straight line and Polyakov loop at nonzero temperature. We discuss the results and conclude in section \ref{sec:discuss}. Some technical details of the calculation have been gathered in Appendix \ref{app:series}.

\section{Straight Wilson line in a CFT}\label{sec:line}

We will start by considering a straight Wilson line in a $d$-dimensional CFT, extended along a spatial direction. The dual is a string in $AdS_{d+1}$ space (and localized in the internal directions). The metric in Poincar\'e coordinates is
\begin{equation}
ds^2=G_{MN}dx^M dx^N=\frac{L^2}{z^2}\left( dz^2+\eta_{\mu\nu}dx^\mu dx^\nu\right).
\end{equation} 
We are using indices $M=\mu,z$ and $\mu=0,1,\dots, d-1$. In the following we will work with dimensionless coordinates by doing the rescaling
\begin{equation}
z\to Lz, \  \ x^\mu\to L x^\mu.
\end{equation}
To specify the action of the string we introduce $\sigma^a=(\tau,\sigma)$ as the (dimensionless) worldsheet coordinates of the string, $X^M(\tau,\sigma)$ as the embedding functions and $h_{ab}$ as the worldsheet metric. Removing the overall $L^2$ factor, the induced metric is
\begin{equation}
g_{ab}=\frac{1}{Z^2}\eta_{MN}\partial_a X^M\partial_b X^N.
\end{equation}
Then, the Polyakov action for a string of tension $T_s$ is
\begin{equation}
S_P=\frac{T_sL^2}{2}\int d^2 \sigma \sqrt{h} h^{ab} g_{ab}+\phi_0 \chi_E.
\end{equation}
Where $\chi_E$ is the Euler characteristic of the string surface with a coefficient proportional to the constant dilaton $\phi_0=\log g_s$, with $g_s$ the string coupling. If boundary terms are properly accounted for, the Euler characteristic is just a constant determined by the string topology.

The Polyakov action is invariant under both worldsheet diffeomorphisms and Weyl transformations, which are gauge symmetries of the string. We will use them to fix the metric to be (Euclidean) $AdS_2$, in the Poincar\'e patch
\begin{equation}
h_{ab}=\frac{1}{\sigma^2}\delta_{ab}.
\end{equation}
After the gauge fixing there is a remnant conformal symmetry that leaves the metric invariant and consists of simultaneous worldsheet diffeomorphisms and Weyl transformations. This is a true symmetry of the string that corresponds to the reparametrization invariance of the Wilson loop, as it will be clear later. 

For a straight Wilson line along the $x^1$ direction we need to impose boundary conditions on the string. By our choice of metric $\sigma=0$ should correspond to the boundary of the worldsheet, so that
\begin{equation}
\lim_{\sigma\to 0} X^1=x^1,\ \ \lim_{\sigma\to 0} X^M=0,\; M\neq 0.
\end{equation}
As we have formulated them, these conditions are invariant under conformal transformations. To ease the notation in the following we will use $X=X^1$ and $Z=X^z$. 

We have some restrictions on the embedding functions. In the first place, the induced metric should be compatible with the worldsheet metric (the worldsheet energy-momentum tensor vanishes)
\begin{equation}\label{eq:metricconst}
g_{ab}-\frac{1}{2}h_{ac} h^{bd}g_{cd}=0.
\end{equation}
And in the second place the embedding functions have to satisfy the equations of motion
\begin{equation}\label{eq:stringeoms}
\frac{1}{\sqrt{h}}\partial_a\left( \sqrt{h} h^{ab}\frac{\partial_b X^M}{Z^2}\right)+\frac{2}{Z}h^{ab}g_{ab}\delta^M_z=0.
\end{equation}
The simplest solution is a string extended along the $(x^1,z)$ directions
\begin{equation}\label{eq:simplesol}
X=\tau, \ \ Z=\sigma,\ \ X^M=0, M\neq 1,z.
\end{equation}

The area of the string is divergent, it can be regularized by introducing appropriate local counterterms at a cutoff $z=\epsilon$ that will eventually be sent to the boundary $\epsilon\to 0$. The regularized action is
\begin{equation}
S_\epsilon=\frac{T_sL^2}{2}\int_{\sigma>\epsilon} d^2 \sigma \sqrt{h} h^{ab} g_{ab}-T_s L^2\int_{\sigma=\epsilon} d\tau\, e_\tau +\phi_0\chi_E.
\end{equation}
Where  $e_\tau=\sqrt{h_{\tau\tau}}$ is the einbein. The Euler characteristic is
\begin{equation}\label{eq:chiE}
\chi_E=\frac{1}{4\pi}\left[\int_{\sigma>\epsilon} d^2\sigma \sqrt{h}R+2\int_{\sigma=\epsilon} d\tau e_\tau K\right]=1,
\end{equation}
where $R$ is the Ricci scalar of the worldsheet metric and $K$ is the extrinsic curvature.

\subsection{String solution with arbitrary boundary reparametrizations}

Instead of the simple solution \eqref{eq:simplesol} we may consider an arbitrary reparametrization of the line at the boundary
\begin{equation}
\lim_{\sigma\to 0} X=x_0(\tau),
\end{equation}
without modifying the shape of the string in the embedding space. This implies modifying the embedding functions $X=X(\tau,\sigma)$, $Z=Z(\tau,\sigma)$ and keeping $X^M=0$ for $M\neq 1,z$. The constraint \eqref{eq:metricconst} can be satisfied as long as the induced metric is conformally flat $g_{ab}=\Omega \delta_{ab}$. From now on, let us denote $\partial_\tau=\dot{}$, $\partial_\sigma={}'$. The induced metric for this more general embedding is
\begin{equation}\label{eq:inducedline}
g_{ab}=\frac{1}{Z^2}\left(\begin{array}{cc} \dot{X}^2+\dot{Z}^2 & \dot{X}X'+\dot{Z}Z' \\ \dot{X}X'+\dot{Z}Z' & (X')^2+(Z')^2 \end{array}\right). 
\end{equation}
In order to have a conformally flat metric, the conditions we need to impose are
\begin{equation}\label{eq:eqsconfflat}
\dot{X}X'+\dot{Z}Z'=0,\ \  \dot{X}^2+\dot{Z}^2=(X')^2+(Z')^2.
\end{equation}
Which can be solved by
\begin{equation}\label{eq:eqsconfflat2}
Z'=\dot{X},\ \ X'=-\dot{Z},\ \ X''+\ddot{X}=0, \ \ Z''+\ddot{Z}=0.
\end{equation}
It can be easily check that solutions to the equations above are also solutions to the equations of motion \eqref{eq:stringeoms}.

The solutions are
\begin{equation}
\begin{split}\label{eq:XZline}
X=\int_{-\infty}^\infty d\tau_0 \frac{1}{\pi}\frac{\sigma}{\sigma^2+(\tau-\tau_0)^2} x_0(\tau_0),\\
Z=\int_{-\infty}^\infty d\tau_0 \frac{1}{\pi}\frac{\tau-\tau_0}{\sigma^2+(\tau-\tau_0)^2} x_0(\tau_0).
\end{split}
\end{equation}
However, when the derivatives of $x_0$ are small compared to $1/\sigma$, it is more interesting to express the solutions as an infinite series expansion (see Appendix~\ref{app:series})
\begin{equation}
\begin{split}\label{eq:XZexpline}
X=&\cos\left(\sigma\frac{d}{d\tau}\right) x_0(\tau)=x_0-\frac{1}{2}\sigma^2 \ddot{x}_0+\frac{1}{24}\sigma^4 x_0^{(4)}+\cdots,\\
Z=&\sin\left(\sigma\frac{d}{d\tau}\right) x_0(\tau)=\sigma \dot{x}_0-\frac{1}{6}\sigma^3 \dddot{x}_0+\frac{1}{120}\sigma^5 x_0^{(5)}+\cdots.
\end{split}
\end{equation}
In this form it is straightforward to find the conformal factor in the induced metric in a similar expansion
\begin{equation}
\Omega=\frac{1}{\sigma^2}-\frac{2}{3}\{ x_0,\tau\}+\sigma^2\left(\frac{1}{15}\partial_\tau^2\{x_0,\tau\}+\frac{4}{15}(\{x_0,\tau\})^2 \right)+\cdots.
\end{equation}
Here we have introduced the Schwarzian derivative
\begin{equation}
\{ x_0,\tau\}=\frac{\dddot{x}_0}{\dot{x}_0}-\frac{3}{2}\left(\frac{\ddot{x}_0}{\dot{x}_0} \right)^2.
\end{equation}
Higher order terms can also be written in terms of the Schwarzian and its derivatives. The Schwarzian is invariant under $GL(2,\mathbb{R})$ reparametrizations of the form
\begin{equation}\label{eq:trivialrep}
x_0(\tau)\longrightarrow \frac{a x_0+b}{cx_0+d}, \ \ a,b,c,d\in \mathbb{R},\ \ ad-bc\neq 0.
\end{equation}
For $x_0(\tau)=\tau$ the Schwarzian vanishes, these are the transformations induced at the boundary by $AdS_2$ isometries. Defining the complex coordinate $\zeta=\tau+i\sigma$, the $AdS_2$ metric in these coordinates is
\begin{equation}
ds^2=-\frac{4d\zeta d\overline{\zeta}}{(\zeta-\overline{\zeta})^2},
\end{equation}
which is manifestly invariant under the transformation
\begin{equation}
\zeta\longrightarrow \frac{a\zeta+b}{c\zeta+d}.
\end{equation}
When $\sigma\to 0$, $\zeta=\overline{\zeta}=\tau$ leading to the transformations we wrote above. Thus boundary reparametrizations of the form \eqref{eq:trivialrep} with $x_0=\tau$ do not lead to changes in the induced metric and the conformal factor stays fixed as $\Omega=1/\sigma^2$.

Let us show now that a conformal transformation trivializes the embedding. First, we perform a worldsheet diffeomorphism
\begin{equation}
\tau=\tau(\bar{\tau},\bar{\sigma}), \ \ \sigma=\sigma(\bar{\tau},\bar{\sigma}),
\end{equation}
such that
\begin{equation}\label{eq:diffeotrivial}
X(\tau,\sigma)=\bar{\tau},\ \ Z(\tau,\sigma)=\bar{\sigma}.
\end{equation}
In the near boundary expansion the transformed coordinates have expansions similar to $X$ and $Z$
\begin{equation}\label{eq:wsdiffeo}
\tau=t(\bar{\tau})-\frac{1}{2}\ddot{t}(\bar{\tau})\bar{\sigma}^2+\cdots,\ \ \sigma=\dot{t}(\bar{\tau})\bar{\sigma}-\frac{1}{6}\dddot{t}(\bar{\tau})\bar{\sigma}^3+\cdots.
\end{equation}
The induced and worldsheet metrics in the new coordinates are
\begin{equation}
\bar{g}_{ab}=\frac{1}{\bar{\sigma}^2}\delta_{ab},\ \ \bar{h}_{ab}=\bar{\Omega}\delta_{ab},
\end{equation}
where the conformal factor in the worldsheet metric equals to
\begin{equation}\label{eq:wsconffact}
\bar{\Omega}=\frac{1}{\bar{\sigma}^2}-\frac{2}{3}\{ t(\bar{\tau}),\bar{\tau}\}+\cdots.
\end{equation}
Finally, to put back the worlsheet metric in its original form we do a Weyl transformation
\begin{equation}\label{eq:weyltriv}
\bar{h}_{ab}\ \longrightarrow \ h_{ab}=\frac{1}{\bar{\sigma}^2 \bar{\Omega}}\bar{h}_{ab}=\frac{1}{\bar{\sigma}^2}\delta_{ab}.
\end{equation}
This shows that conformal transformations on the worldsheet correspond to reparametrizations of the Wilson loop, as we could in principle follow these steps backwards to produce an arbitrary reparametrization from the trivial embedding.

\subsection{Induced anomalies in the cutoff action}\label{sec:repanomaly}

The expectation value of the Wilson line in the dual field theory is determined by the string action on-shell. We can introduce a cutoff in the radial direction that splits the dual geometry in two parts. The region between the $AdS$ boundary and the radial cutoff is identified with UV degrees of freedom of the field theory dual and the region beyond the cutoff captures the IR degrees of freedom.

The IR region of the geometry still describes the dual of a strongly coupled theory, with a line defect that has a holographic description as a string ending at the cutoff along a line in the $x^1$ direction. In addition to the string, that captures the dynamics of the IR degrees of freedom of the strongly coupled field theory dual, there is an effective action at the cutoff for the defect that is obtained integrating over the radial direction the string action between the boundary and the cutoff. The natural interpretation of the cutoff action is that it captures the effect of the UV degrees of freedom close to the Wilson line after they have been integrated out.

There are two possible natural choices for the cutoff, we could introduce a cutoff in the worldsheet coordinate $\sigma=1/(L\Lambda)$, or we could introduce a cutoff in the radial coordinate of the geometry $z=1/(L\Lambda)$, with $\Lambda$ an energy scale. If the cutoff is taken in the worldsheet, the cutoff action is
\begin{equation}\label{eq:Ssigmacutoff}
S_\Lambda=T_s L^2\int d\tau \left( -L\Lambda-\frac{2}{3}\frac{1}{L\Lambda} \{x_0,\tau\}+\frac{1}{3}\frac{1}{(L\Lambda)^3}\left(\frac{1}{15}\partial_\tau^2\{x_0,\tau\}+\frac{2}{5}(\{x_0,\tau\})^2 \right)+\cdots \right)+\frac{\phi_0}{2\pi}\int d\tau L\Lambda. 
\end{equation}
This shows that the cutoff action is not invariant under reparametrizations. From the bulk perspective, the string extended beyond the cutoff is reparametrization invariant up to boundary terms. This non-invariance is compensated by the action at the cutoff, so that the total action consisting of string plus defect is invariant. This can be seen as analogous to the anomaly inflow between a Chern-Simons action for a gauge field in $2+1$ dimensions and chiral edge modes at a boundary. Thus, we can see the terms depending on the Schwarzian as originating from a reparametrization anomaly in the cutoff action. 

However, one might object that a radial cutoff in the geometry is more natural than a cutoff in the worldsheet, since the $AdS$ radial direction is typically more readily identified with energy scales in the field theory dual. If we fix the radial cutoff, then we should integrate the string action up to a value of the worldsheet coordinate determined by the condition
\begin{equation}
Z(\tau,\sigma_\Lambda(\tau))=1/(L\Lambda).
\end{equation}
The solution can be expanded as
\begin{equation}
\sigma_\Lambda(\tau)=\frac{1}{L \Lambda \dot{x}_0}\left[1+\frac{1}{6} \frac{1}{(L\Lambda)^2} \frac{\dot{x}_0}{(\dot{x}_0)^3}+\frac{1}{6}\frac{1}{(L\Lambda)^4}\frac{10(\ddot{x}_0)^2-\dot{x}_0 x_0^{(5)}}{(\dot{x}_0)^6}+\cdots \right].
\end{equation}
Then, the action integrated up to this value is
\begin{equation}
S_\Lambda=T_s L^2 \int d\tau\left(-L\Lambda \dot{x}_\Lambda\right)+\frac{\phi_0}{2\pi}\int d\tau \frac{1}{\sigma_\Lambda}.
\end{equation}
Where we have defined
\begin{equation}
x_\Lambda=x_0+\frac{1}{2}\frac{1}{(L\Lambda)^2}\frac{\ddot{x}_0}{\dot{x}_0}+\frac{1}{72}\frac{1}{(L\Lambda)^4}\frac{4 \ddot{x}_0\dddot{x}_0-\dot{x}_0 x_0^{(4)}}{(\dot{x}_0)^5}+\cdots .
\end{equation}
Therefore, with this choice of cutoff, the contribution of the induced metric to the defect action is simply a reparametrization of the worldline coordinate $d\tau_\Lambda=d\tau \dot{x}_\Lambda$. This is to be expected because the area of the string between the $AdS$ boundary and the radial cutoff in the geometry should be independent of the reparametrization. However, we must now pay attention to the contribution to the defect action deriving from the Ricci scalar of the worldsheet metric, which now gives a non-trivial contribution
\begin{equation}\label{eq:SLambda}
S_\Lambda=T_s L^2 \int d\tau_\Lambda\left(-L\Lambda \right)+\frac{\phi_0}{2\pi}\int d\tau_\Lambda \left( L\Lambda+\frac{2}{3}\frac{1}{L\Lambda}\{ t(\tau_\Lambda),\tau_\Lambda \}+\cdots\right).
\end{equation}
Where we have defined $t$ as the inverse of $x_0$: $x_0[t(\theta)]=\theta$ and used that
\begin{equation}
\{ t(\theta),\theta \} =-\frac{1}{\dot{x}_0^2}\{ x_0(\tau),\tau\}.
\end{equation}
We can recover the same result by performing the worldsheet diffeomorphism \eqref{eq:diffeotrivial}. Once we have trivialized the embedding, the terms proportional to the induced metric that contribute to the cutoff action are trivial. However, the integral over the Ricci scalar introduce a boundary term proportional to the extrinsic curvature
\begin{equation}
\bar{K}=-\frac{1}{2}\frac{\bar{\Omega}'}{\bar{\Omega}^{3/2}}=1+\bar{\sigma}^2 \{ t(\bar{\tau}),\bar{\tau}\}+\cdots.
\end{equation}
Where $\bar{\Omega}$ is given in \eqref{eq:wsconffact}. The resulting cutoff action is the same we found before \eqref{eq:SLambda} identifying $\tau_\Lambda=\bar{\tau}$
\begin{equation}\label{eq:SZcutoff}
S_\Lambda=T_s L^2\int d\bar{\tau} \left( -L\Lambda \right)+\frac{\phi_0}{2\pi}\int d\bar{\tau} \left(L\Lambda+\frac{2}{3}\frac{1}{L\Lambda}\{ t(\bar{\tau}),\bar{\tau}\}+\cdots\right). 
\end{equation}
The Schwarzian derivative indicates that effectively there is an anomaly at the cutoff, which compensates the non-invariance of the string under worldsheet diffeomorphisms. Note that the Weyl transformation \eqref{eq:weyltriv} would remove this term, so there is another associated anomaly at the cutoff, in such a way that the anomalous terms cancel out for conformal transformations of the worldsheet.

\section{Generalizations}\label{sec:gener}

In order to highlight the universality of the reparametrization anomaly we will now study three straightforward generalizations of spatial Wilson loops in a CFT: a straight Wilson line at nonzero temperature, a circular Wilson line at zero temperature and a Polyakov loop at nonzero temperature.

\subsection{Straight Wilson line at nonzero temperature}

At nonzero temperature the holographic dual of a CFT${}_d$ is an $AdS_{d+1}$ black brane solution, that in Poincar\'e patch coordinates reads
\begin{equation}\label{eq:adsBB}
ds^2=\frac{L^2}{z^2}\left( \frac{dz^2}{f(z)}-f(z)(dx^0)^2+\delta_{ij}dx^i dx^j\right), \ \ f(z)=1-\left(\frac{z}{z_H}\right)^d.
\end{equation}
The temperature of the dual CFT is $T=\frac{d}{4\pi z_H}$. It will be convenient for us to do a change of coordinates such that the induced metric on the string becomes conformally flat. This can be achieved by picking a new radial coordinate $u$ such that
\begin{equation}
du=\frac{dz}{\sqrt{f(z)}}.
\end{equation}
The solution is
\begin{equation}\label{eq:usol1}
u=\frac{z_H
   B_{\left(\frac{z}{z_H}\right)^d}\left(\frac{1}{d},\frac{1}{2}\right)}{d}.
\end{equation}
Where $B_x(a,b)$ is the incomplete Beta function. The horizon in the $u$ coordinate is located at
\begin{equation}
u_H=\frac{B\left(\frac{1}{d},\frac{1}{2}\right)}{d}z_H
\end{equation}
The relation can be inverted to
\begin{equation}
\left(\frac{z(u)}{z_H}\right)^d= I^{-1}_{\frac{u}{u_H}}\left(\frac{1}{d},\frac{1}{2}\right),
\end{equation}
where $I_x(a,b)=B_x(a,b)/B(a,b)$ is the regularized incomplete Beta function and $I^{-1}_x(a,b)$ is its inverse. For convenience let us do the following rescaling of the coordinates
\begin{equation}
u\to u_H u,\ \ z\to z_H z,\ \ x^\mu\to u_H x^\mu.
\end{equation}
Then, the metric is
\begin{equation}
ds^2=\frac{\tilde{L}^2}{z(u)^2}\left(du^2-f[z(u)](dx^0)^2+\delta_{ij}dx^i dx^j\right),\ \ f(z)=1-z^d,\ \ z(u)^d=I^{-1}_u\left(\frac{1}{d},\frac{1}{2}\right),
\end{equation}
where $\tilde{L}=L u_H/z_H=B\left(\frac{1}{d},\frac{1}{2}\right) L/d$.

The worldsheet action dual to a Wilson line in the black brane geometry is
\begin{equation}
S_P=\frac{T_s\tilde{L}^2}{2}\int_{u<1} d^2 \sigma \sqrt{h} h^{ab} g_{ab}+\phi_0 \hat{\chi}_E.
\end{equation}
If the string reaches the black brane horizon, only the part of the string outside the horizon is taken into account. This introduces a cutoff in the radial direction at $u=1$. The term $\hat{\chi}_E$ equals \eqref{eq:chiE} with the same cutoff at $u=1$, but without a extrinsic curvature term at the horizon, so it is no longer equal to the Euler characteristic of the string worldsheet and does not take integer values in general.

For a straight spatial Wilson line, we can take as embeddding and worldsheet metric
\begin{equation}
X^1\equiv X=\tau, \ \ X^u\equiv U=\sigma,\ \ X^M=0, M\neq 1,u.\ \ h_{ab}=\frac{1}{\sigma^2}\delta_{ab},
\end{equation}
With this choice the induced metric is the same as the worldsheet metric up an overall factor, which automatically satisfies the constraint \eqref{eq:metricconst}. The equations for the embedding functions, which now have the form
\begin{equation}\label{eq:stringeoms2}
\frac{1}{\sqrt{h}}\partial_a\left( \sqrt{h} h^{ab}\frac{\partial_b X^M}{z(U)^2}\right)+\frac{2}{z(U)}z'(U)h^{ab}g_{ab}\delta^M_u=0,
\end{equation}
are also satisfied.

As in the zero temperature case we consider an arbitrary reparametrization of the Wilson line at the boundary
\begin{equation}
\lim_{\sigma\to 0} X=x_0(\tau).
\end{equation}
The embedding functions will be modified as before $X(\tau,\sigma)$ and $U(\tau,\sigma)$. The induced string metric is
\begin{equation}
g_{ab}=\frac{1}{z(U)^2}\left(\begin{array}{cc} \dot{X}^2+\dot{U}^2 & \dot{X}X'+\dot{U}U' \\ \dot{X}X'+\dot{U}U' & (X')^2+(U')^2 \end{array}\right),
\end{equation}
which can be made conformally flat by imposing the same conditions as at zero temperature \eqref{eq:eqsconfflat} and \eqref{eq:eqsconfflat2}, simply replacing $Z$ by $U$. As happened at zero temperature the embedding equations of motion \eqref{eq:stringeoms2} are automatically satisfied even with the new conformal factor.

Following the discussion at zero temperature, we can fix the radial cutoff in the geometry, but now on the $u$ coordinate
\begin{equation}
U\left(\tau,\sigma_\Lambda(\tau)\right)=1/(\Lambda L).
\end{equation}
Performing the same worldsheet diffeomorphism as before \eqref{eq:wsdiffeo}, the induced and worldsheet metric become
\begin{equation}
g_{ab}=\frac{1}{z(\bar{\sigma})^2} \delta_{ab},\ \ h_{ab}=\bar{\Omega}\delta_{ab},
\end{equation}
where the conformal factor is the same as at zero temperature \eqref{eq:wsconffact}. 

It follows that the cutoff effective action is the same at zero and nonzero temperature. However, at nonzero temperature there is a physical cutoff that is the black brane horizon, where $U=1$, implying $\Lambda L=1$, in our variables. Then, the effective  action after integrating all the way to the horizon is
\begin{equation}
S_H=T_s \tilde{L}^2\int d\bar{\tau} \left(\int_\epsilon^1 d\bar{\sigma} \frac{1}{z(\bar{\sigma})^2}-\frac{1}{\epsilon} \right)+\frac{\phi_0}{2\pi}\int d\bar{\tau} \left(1+\frac{2}{3}\{ t(\bar{\tau}),\bar{\tau}\}+\cdots\right). 
\end{equation}
Restoring units $x_0\to x_0/u_H$ and $\bar{\tau}\to \bar{\tau}/u_H$, the Schwarzian term is
\begin{equation}\label{eq:effectiveT}
S_{Sch}=\frac{\phi_0}{12\pi^2} \frac{B\left(\frac{1}{d},\frac{1}{2} \right)}{T}\int d\bar{\tau} \, \{ t(\bar{\tau}),\bar{\tau}\}
\end{equation}

\subsection{Circular Wilson loop}

We will consider now a Wilson loop defined on a circle of radius $r_0$ localized on some plane, that we can take to be along the $x^\mu$, $\mu=1,2$ directions without loss of generality. In this case it is more convenient to work with polar coordinates in the plane. The metric in the Poincaré patch is:
\begin{equation}
ds^2 = \frac{L^2}{z^2}\left(dz^2+ d r^2 +r^2 d\theta^2+\sum_{\mu=3}^{d-1} (d x^\mu)^2\right).
\end{equation}
The solution for a string ending on a circle of radius $R$ on the boundary lies on the spherical surface \cite{Berenstein:1998ij,Drukker:1999zq}
\begin{equation}
z^2+r^2=r_0^2.
\end{equation}
This surface can be parametrized by the worldsheet embedding
\begin{equation}
X^\theta\equiv \Theta=\tau,\ \ X^r\equiv R=\frac{r_0}{\cosh \sigma},\ \ Z=r_0\tanh \sigma, \ \ X^M=0,\ M\neq r,\theta,z.
\end{equation}
Where both $\tau$ and $\theta$ have $2\pi$ periodicity. 

This yields global $AdS_2$ in conformally flat coordinates as the induced metric on the string
\begin{equation}\label{eq:circmetric}
ds_2^2 =g_{ab}d\sigma^a d\sigma^b= \frac{1}{\sinh^2 \sigma}\left(d\tau^2+d\sigma^2\right).
\end{equation}
Given the topology of the string worldsheet, we will select the string metric to be the same
\begin{equation}
h_{ab}=\frac{1}{\sinh^2\sigma}\delta_{ab}.
\end{equation}
Let us now consider a general reparametrization of the embedding of the form
\begin{equation}\label{eq:embedcircle}
\Theta=q \tau+\theta(\tau,\sigma),\ \ R=\frac{r_0}{\cosh S},\ \ Z=r_0\tanh S,\ \ S=q\sigma+ s(\tau,\sigma).
\end{equation}
Where the periodicity of $\tau$ is now $2\pi p$ and $p,q$ are nonzero integers. Both $\theta$ and $s$ are taken to be periodic functions of $\tau$
\begin{equation}
\theta(\sigma,\tau+2\pi p)=\theta(\sigma,\tau),\ \ s(\sigma,\tau+2\pi p)=s(\sigma,\tau).
\end{equation}
At the $AdS$ boundary $\sigma=0$ we impose the conditions
\begin{equation}
\Theta(\sigma=0,\tau) =\Theta_0(\tau)=q\tau+\theta_0(\tau),\ \ S(\sigma=0,\tau)=0,\ \ \theta_0(\tau+2\pi p)=\theta_0(\tau).
\end{equation}
The term in $\Theta_0$ that is linear in $\tau$ indicates that the Wilson line is winding $w=pq$ times over the circle. However, unless $q=1/p$ (so $w=1$) the induced metric will have a conical singularity. We will ignore this and proceed with general values of $p$, $q$.

With this embedding, the induced metric and worldsheet metric become
\begin{equation}
g_{ab}=\frac{1}{\sinh^2 S}\left(\begin{array}{cc} \dot{S}^2+\dot{\Theta}^2 & \dot{S}S'+\dot{\Theta}\Theta' \\ \dot{S}S'+\dot{\Theta}\Theta' & (S')^2+(\Theta')^2 \end{array}\right),\ \ h_{ab}=\frac{q^2}{\sinh^2(q\sigma)}\delta_{ab}. 
\end{equation}
This takes the same form as for the straight line \eqref{eq:inducedline}, so the induced metric can be made conformally flat  for embedding solutions satisfying the same set of equations as given in \eqref{eq:eqsconfflat2}
\begin{equation}
S'=\dot{\Theta},\ \ \Theta'=-\dot{S},\ \ S''+\ddot{S}=0, \ \ \Theta''+\ddot{\Theta}=0.
\end{equation}
The linear terms proportional to $q$ in the embedding functions \eqref{eq:embedcircle} automatically satisfy these equations. We can give a solution generalizing the straight line results to account for the periodicity of $\tau$. First we define the functions
\begin{equation}
\begin{split}
G_\Theta(\sigma,\tau-\tau_0)=&\sum_{n=-\infty}^\infty \frac{1}{\pi} \frac{\sigma}{\sigma^2+(\tau-\tau_0+2\pi n p)^2}=\frac{1}{2\pi p}\frac{\sinh\frac{\sigma}{p}}{\cosh\frac{\sigma}{p}-\cos\frac{\tau-\tau_0}{p}},\\
G_S(\sigma,\tau-\tau_0)=&\frac{1}{2\pi p}\frac{\sin\frac{\tau-\tau_0}{p}}{\cosh\frac{\sigma}{p}-\cos\frac{\tau-\tau_0}{p}}.
\end{split}
\end{equation}
Then, the solutions for the embedding functions are
\begin{equation}\label{eq:XZcircle}
\begin{split}
\theta(\sigma,\tau)=&\int_{-\pi p}^{\pi p} d\tau_0\, G_\Theta(\sigma,\tau-\tau_0)\theta_0(\tau_0),\\
s(\sigma,\tau)=&\int_{-\pi p}^{\pi p} d\tau_0\, G_S(\sigma,\tau-\tau_0)\theta_0(\tau_0).
\end{split}
\end{equation}
One can recover the straight line expressions \eqref{eq:XZline} by taking the $p\to \infty$ limit.

As before, it will be more convenient for us to use an expansion of the solutions for small $\tau$ derivatives relative to $1/\sigma$, which is actually of the same form as for the straight line \eqref{eq:XZexpline} (see Appendix~\ref{app:series}) 
\begin{equation}\label{eq:XZexpcircle}
\begin{split}
\Theta=&\cos\left(\sigma\frac{d}{d\tau}\right) \Theta_0(\tau)=\Theta_0-\frac{1}{2}\sigma^2 \ddot{\Theta}_0+\frac{1}{24}\sigma^4 \Theta_0^{(4)}+\cdots,\\
S=&\sin\left(\sigma\frac{d}{d\tau}\right) \Theta_0(\tau)=\sigma \dot{\Theta}_0-\frac{1}{6}\sigma^3 \dddot{\Theta}_0+\frac{1}{120}\sigma^5 \Theta_0^{(5)}+\cdots.
\end{split}
\end{equation}
However, before doing the reparametrization, we end up with a slightly modified result, since the conformal factor in the induced metric  was different:
\begin{equation}
\Omega = \frac{1}{\sigma^2}-\frac{2}{3}\left\{\tan\frac{\Theta_0}{2},\tau\right\}+\sigma^2\left(\frac{1}{15}\partial_\tau^2\left\{\tan\frac{\Theta_0}{2},\tau\right\}+\frac{4}{15}\left\{\tan\frac{\Theta_0}{2},\tau\right\}^2\right)+\cdots.
\end{equation}
Where the Schwarzian terms are now
\begin{equation}
\left\{\tan\frac{\Theta_0}{2},\tau\right\}=\left\{ \Theta_0,\tau\right\}+\frac{1}{2}\dot{\Theta}^2_0.
\end{equation}
In this case the terms that appear in the expansion are invariant under boundary reparametrizations of the form
\begin{equation}\label{eq:su11}
e^{i\Theta_0(\tau)}\longrightarrow \frac{\alpha e^{i\Theta_0}+\bar{\beta}}{\beta e^{i\Theta_0}+\bar{\alpha}},\ \ \alpha,\beta \in \mathbb{C},\  |\alpha|^2-|\beta|^2=1.
\end{equation}
This can be understood as the boundary limit of the $SU(1,1)$ isometry transformations of the global $AdS_2$ metric. The symmetry is more easily realized in the coordinates
\begin{equation}
\cosh \sigma=\frac{1}{\tanh\rho};\;\;\;\zeta=\tanh\frac{\rho}{2}e^{i\tau}, \;\bar{\zeta}=\tanh\frac{\rho}{2}e^{-i\tau},
\end{equation}
leading to the metric
\begin{equation}
ds_2^2=\frac{4 d\zeta d\bar{\zeta}}{(1-|\zeta|^2)^2}.
\end{equation}
The $SU(1,1)$ isometry transformations in these coordinates are
\begin{equation}
\zeta\longrightarrow\frac{\alpha\zeta+\bar{\beta}}{\beta\zeta+\bar{\alpha}} \underset{|\zeta|\to 1}{\longrightarrow} \frac{\alpha e^{i\tau}+\bar{\beta}}{\beta e^{i\tau}+\bar{\alpha}}.
\end{equation}
Thus leading to the transformations \eqref{eq:su11}. However, contrary to the straight line, the Schwarzian derivative is nonzero for the trivial embedding, rather for $\Theta_0=\tau$ or any $SU(1,1)$ equivalent,
\begin{equation}
\left\{\tan\frac{\Theta_0}{2},\tau\right\}=\frac{1}{2}.
\end{equation}
This is of course necessary in order to recover the expansion of the conformal factor
\begin{equation}\label{eq:CFcircle}
\Omega=\frac{1}{\sinh^2\sigma}= \frac{1}{\sigma^2}-\frac{1}{3}+\frac{\sigma^2}{15}+\cdots.
\end{equation}

Aside from the difference on the isometry group of the induced metric and the corresponding invariant Schwarzian derivatives, the analysis of the straight Wilson line of section \ref{sec:repanomaly} can be generalized without any other modifications to the circular Wilson loop. Therefore, replacing
\begin{equation}
\left\{x_0,\tau\right\}\to \left\{ \tan\frac{\Theta_0}{2},\tau\right\},
\end{equation}
the cutoff action takes the form \eqref{eq:Ssigmacutoff} if we introduce the cutoff in the worldsheet coordinates $\sigma=1/(L\Lambda)$. If the cutoff is in the radial coordinate of the background $z=1/(L\Lambda)$, then the cutoff action is \eqref{eq:SZcutoff} changing
\begin{equation}\label{eq:schwzcircle}
\left\{t(\bar{\tau}),\bar{\tau}\right\}\to \left\{ \tan\frac{t(\bar{\tau})}{2},\bar{\tau}\right\}.
\end{equation}
Where $\bar{\sigma}=S(\sigma,\tau)$, $\bar{\tau}=\Theta(\sigma,\tau)$ define the barred coordinates and $t(\bar{\tau})$ is the inverse of the boundary reparametrization $\Theta_0[t(\theta)]=\theta$.

\subsection{Polyakov Loop}

At finite temperature a Polyakov loop is defined as a Wilson line wrapping the time direction after a Wick rotation to Euclidean signature. A nonzero expectation value for the Polyakov loop implies a spontaneous breaking of center symmetry and it is taken as an indication of deconfinement in pure Yang-Mills. The holographic dual is a string wrapped around the Euclidean time direction of the Wick rotated $AdS_{d+1}$ black brane \eqref{eq:adsBB}, with metric
\begin{equation}
ds^2=\frac{L^2}{z^2}\left( \frac{dz^2}{f(z)}+f(z)dt_E^2+\delta_{ij}dx^i dx^j\right), \ \ f(z)=1-\left(\frac{z}{z_H}\right)^d.
\end{equation}
Where the Euclidean time direction $t_E$ has periodicity $\beta=1/T$. The string dual to the Polyakov loop will be extended along the $(z,t_E)$ directions, and has the topology of a disk. In order to have a conformally flat induced metric we first rescale all the coordinates $x^M\to z_H x^M$ and then introduce the following change of variables for the radial coordinate
\begin{equation}
du=\frac{dz}{f(z)}.
\end{equation}
The solution to this equation is similar to the one we found for a spatial Wilson loop \eqref{eq:usol1}
\begin{equation}
u = \frac{B_{z^d}\left(\frac{1}{d},0\right)}{d}.
\end{equation}
In this case the radial coordinate is not bounded, close to the horizon $z\to 1$
\begin{equation}\label{eq:uHpol}
u\sim -\frac{1}{d}\log(1-z)\to +\infty.
\end{equation}
We now follow similar steps, and rescale the coordinates as follows
\begin{equation}
u= \frac{1}{2\pi z_H T} r,\ \ \ \ t_E= \frac{1}{2\pi z_H T}\theta.
\end{equation}
After this, the periodicity of $\theta$ is $2 \pi$, and the metric takes the form
\begin{equation}
ds^2=\frac{L^2}{z(r)^2}\left(\frac{2}{d}\right)^2f\left[z(r)\right]\left(dr^2+d\theta^2\right)+\frac{L^2}{z(r)^2}\delta_{ij}dx^i dx^j,\ \ f(z)=1-z^d,\ \ z(r)^d=B^{-1}_{2r}\left(\frac{1}{d},0\right),
\end{equation}
where $B^{-1}_z(x,y)$ is the inverse of the incomplete beta function. The simplest choice for the string embedding of the Polyakov loop is
\begin{equation}
X^\theta\equiv \Theta = \tau, \ \ X^r\equiv R =\sigma, \ \ X^i=0,
\end{equation}
where $\tau$ is periodic, with periodicity $2\pi$. The induced metric is, after removing an overall $L$ factor,
\begin{equation}
ds_2^2=g_{ab}dx^a dx^b=\left(\frac{2}{d}\right)^2\frac{f[z(\sigma)]}{z(\sigma)^2}(d\tau^2+d\sigma^2).
\end{equation}
This is the same type of induced metric we found for the circular Wilson loop \eqref{eq:circmetric}. The asymptotic behavior for $\sigma\to 0$ is that of $AdS_2$, with the conformal factor $\sim 1/\sigma^2$. For $\sigma\to \infty$ we see from \eqref{eq:uHpol} that $z^d\simeq 1-e^{-2 \sigma}$, in such a way that the conformal factor is
\begin{equation}
\left(\frac{2}{d}\right)^2\frac{f[z(\sigma)]}{z(\sigma)^2}\sim 4e^{-2\sigma}\sim \frac{1}{\sinh^2\sigma},
\end{equation}
which is the expected behavior for $AdS_2$. We are thus driven to take as worldsheet metric corresponding to global $AdS_2$
\begin{equation}
h_{ab}=\frac{1}{\sinh^2\sigma}\delta_{ab}.
\end{equation}
From this point onwards we can proceed following the same steps as for a circular Wilson loop, introducing a reparametrization that is the analog of \eqref{eq:embedcircle}
\begin{equation}
\Theta=q \tau+\theta(\tau,\sigma),\ \ R=S=q\sigma+ s(\tau,\sigma).
\end{equation}
The details for the solutions and the symmetries of the worldsheet metric are the same as for the circular Wilson loop, so we arrive to the same result of a Schwarzian action for the worldsheet diffeomorphisms \eqref{eq:schwzcircle}.

\section{Discussion}\label{sec:discuss}

Our motivation was to study the low energy effective description of a Wilson loop using the gauge/gravity duality, by taking a string in some IR region of the dual geometry determined by a cutoff in the holographic radial direction. First, we have identified a map between reparametrizations of the Wilson loop and conformal transformations in the worldsheet of the dual string. We have then shown that the string with a cutoff in the worldsheet is not invariant under reparametrizations of the Wilson loop, so that it is necessary to add a cutoff action proportional to the Schwarzian derivative of said reparametrizations, as well as higher derivative terms further suppressed by the cutoff scale. On the other hand, if the cutoff is set in the target space, the string is invariant under Wilson loop reparametrizations, but not under worldsheet diffeomorphisms or Weyl transformations that do not belong to the subset of conformal transformations. Therefore, new terms for the Polyakov string action should be added at the cutoff before fixing the gauge.

In the case of a cutoff on the worldsheet, the new terms have the appropriate structure to describe the effective theory of Goldstone bosons for a spontaneously broken reparametrization invariance of the Wilson loop, similarly to nearly-$AdS_2$ dynamics. They are necessary because a conformal transformation on the worldsheet would change the physical location of the cutoff in the target space, thus modifying the area of the string in the IR region. This is compensated by the change in the cutoff action, so the total area defined as the string action including the cutoff remains the same. Although we have only considered strings in $AdS$ spacetimes, in principle conformal invariance of the worldsheet should be maintained in any string background, so we expect reparametrization invariance of Wilson loops to hold in general. It would be interesting to explore if the Schwarzian in the cutoff action is related with maximal chaos as observed in strings with worldsheet horizons \cite{deBoer:2017xdk,Murata:2017rbp}, similarly to SYK \cite{Maldacena:2016hyu} and JT gravity  \cite{Jensen:2016pah}. 

Another interesting extension of this work would be to include in the analysis non-trivial profiles of the string in the transverse directions, generalizing the results of \cite{Casalderrey-Solana:2019vnc,Gutiez:2020sxg}. It should be noted that a Schwarzian action for transverse fluctuations can also appear when the string is embedded in $AdS_3$ \cite{Banerjee:2018kwy,Vegh:2019any}, but it is related to diffeomorphisms in the target space, rather than to worldsheet transformations.

Finally, let us comment further on the connection and differences between the Wilson loop and SYK. A supersymmetric Wilson loop in higher representations can be described as a brane intersection with dynamical fields on the defect \cite{Gomis:2006sb}. A similar defect theory may be expected to describe a Wilson loop in the fundamental representation. For a 1/2 BPS loop, or at lower order in perturbation theory, the defect action of a straight or circular Wilson loop is reparametrization invariant \cite{Hoyos:2018jky} and this reduces the calculation of the expectation value to a random matrix integral. The SYK model could be seen similarly as a 0+1 defect theory \footnote{They are not exactly the same type of theories as the SYK model has disorder as an additional ingredient. However there are other models without disorder that capture similar physics \cite{Witten:2016iux}, so this does not seem to be a crucial ingredient for the physics.}, where reparametrization invariance emerges at low energies. Contrary to the Wilson loop, in this case it is not a true symmetry, and there is a Schwarzian effective action for reparametrizations. However, as the Schwarzian is an irrelevant deformation in the 0+1 theory, it seems natural that the IR is captured by a random matrix theory as the late time analysis of partition functions suggests \cite{Cotler:2016fpe}. Similarly, random matrix models have been shown to determine partition functions in JT gravity \cite{Saad:2019lba}.

\section*{Acknowledgements}
We thank Andy O'Bannon for useful comments. D.G. is partially supported by the Consejer\'{\i}a de Ciencia, Innovaci\'on y Universidad del Principado de Asturias through the ``Severo Ochoa'' fellowship, PA-20-PF-BP19-044. D.G. and C.H. are partially supported by the AEI through the Spanish grant PGC2018-096894-B-100 and by FICYT through the Asturian grant SV-PA-21-AYUD/2021/52177.

\appendix

\section{Series expansions in $\sigma$}\label{app:series}

First we check the values of $X$ and $Z$ at $\sigma=0$ for the straight Wilson line. Taking \eqref{eq:XZline} and doing the change of variables in the integral $\tau_0=\tau+\sigma v$, the expressions for $X$ and $Z$ become
\begin{equation}
\begin{split}
X(\sigma,\tau)=&\int_{-\infty}^\infty dv \frac{1}{\pi}\frac{1}{1+v^2} x_0(\tau+\sigma v),\\
Z(\sigma,\tau)=&\dashint_{-\infty}^\infty  dv \frac{1}{\pi}\frac{v}{1+v^2} x_0(\tau+\sigma v),
\end{split}
\end{equation}
It is immediate to check that $X(\sigma=0,\tau)=x_0(\tau)$ and $Z(\sigma=0,\tau)=0$. From the equations \eqref{eq:eqsconfflat2}, $X$ is an even function of $\sigma$ and $Z$ an odd function. Therefore, they have the expansions
\begin{equation}
\begin{split}
X(\sigma,\tau)=&\sum_{n=0}^\infty \frac{\sigma^{2n}}{(2n)!} \partial^{2n}_\sigma X\Big|_{\sigma=0},\\
Z(\sigma,\tau)=&\sum_{n=0}^\infty \frac{\sigma^{2n+1}}{(2n+1)!} \partial^{2n+1}_\sigma Z\Big|_{\sigma=0}.
\end{split}
\end{equation}
We can use \eqref{eq:eqsconfflat2} to trade $\sigma$ derivatives by $\tau$ derivatives
\begin{equation}
\begin{split}
X(\sigma,\tau)=&\sum_{n=0}^\infty \frac{\sigma^{2n}}{(2n)!} (-1)^n\partial^{2n}_\tau X\Big|_{\sigma=0},\\
Z(\sigma,\tau)=&\sum_{n=0}^\infty \frac{\sigma^{2n+1}}{(2n+1)!} (-1)^{2n}\partial^{2n+1}_\tau X\Big|_{\sigma=0}.
\end{split}
\end{equation}
But, using that the kernel in the integrand of \eqref{eq:XZline}  depends only on $\tau-\tau_0$,
\begin{equation}
\partial_\tau^N X=(-1)^N\int_{-\infty}^\infty d\tau_0\,\partial_{\tau_0}^N\left(\frac{1}{\pi}\frac{\sigma}{\sigma^2+(\tau-\tau_0)^2} \right)  x_0(\tau_0)=\int_{-\infty}^\infty d\tau_0\frac{1}{\pi}\frac{\sigma}{\sigma^2+(\tau-\tau_0)^2}  x_0^{(N)}(\tau_0). 
\end{equation}
Therefore, $\partial_\tau^N X\Big|_{\sigma=0}=x_0^{(N)}(\tau)$, so we arrive at  \eqref{eq:XZexpline}.

For the circle we can proceed in a similar way. Starting with \eqref{eq:XZcircle} and doing the change of variables
\begin{equation}
\tau_0=\tau+2\arctan\left(\tan\frac{\sigma}{2}v \right), \ \ \sigma=\operatorname{arcsinh}(s).
\end{equation}
We arrive at
\begin{equation}
\begin{split}
\Theta(s,\tau)=&q\tau+\int_{-\infty}^\infty dv \frac{1}{\pi}\frac{1}{1+v^2}\theta_0\left(\tau+2\arctan\left(\frac{s v}{1+\sqrt{1+s^2}} \right)
\right),\\
S(s,\tau)=&q\sigma(s)+\dashint_{-\infty}^\infty dv \frac{1}{\pi}\frac{v}{1+v^2}\frac{1+\sqrt{1+s^2}}{1+\sqrt{1+s^2}+\frac{1+v^2}{2}s^2}\theta_0\left(\tau+2\arctan\left(\frac{s v}{1+\sqrt{1+s^2}} \right)\right).
\end{split}
\end{equation}
So, indeed $\Theta(\sigma=0,\tau)=\Theta(s=0,\tau)=\Theta_0(\tau)$ and $S(\sigma=0,\tau)=S(s=0,\tau)=0$. Since $\Theta$ and $S$ for the circle satisfy the same equations as for the straight line, we can apply the same derivation and arrive at the same result for the expansions. 

In both cases the form of the expansion can also be explicitly checked by taking the $\sigma$ derivative of the integrands, performing the same changes of variables in the integrals we have introduced now, and then computing the integrals explicitly.

\bibliographystyle{JHEP}
\bibliography{references}

\end{document}